\documentstyle[aps,epsfig,twocolumn,floats,array,multirow]{revtex}
\newcommand{\ra}{\rangle}

\title{Entanglement and quantum computation with ions in
  thermal motion}

\author{Anders S\o rensen and Klaus M\o lmer\\
{\small
  Institute of Physics and Astronomy, University of Aarhus}\\
{\small DK-8000 \AA rhus C}}
\date{\today}

\begin{document}
\maketitle

\begin{abstract}
With bichromatic fields it is possible to deterministically produce
entangled states of trapped ions. In this paper we present a unified
analysis of this process for both weak and strong fields, for slow and fast
gates. Simple expressions for the fidelity of creating maximally entangled
states of two or an arbitrary number of ions under non-ideal conditions are
derived and discussed.
\end{abstract}

\bigskip

Pacs. 03.67.Lx, 03.65.Bz, 89.70.+c

\bigskip
\section{Introduction}
Quantum computing relies on the ability to perform a collection
of unitary evolutions of a quantum system, 
composed of a number of two-level systems (the qubits), and
it is a key result that a small set of so-called
universal gates exists, which may form the basis
for the entire computation \cite{gates}. 
The development of proposals for physical implementation of 
quantum computing have followed different routes,
according to the various views one may have on the quantum 
dynamical processes. (i): one may view a gate operation 
on a single or on several qubits as a
controlled transition from the initial to the final states,
and one may implement it by a Hamiltonian, or a sequence of Hamiltonians,
that couple 
these states directly. (ii): one may consider
Hamiltonians that couple quite many states, but where 
unwanted operations are dynamically
suppressed by resonance conditions or by 'bang-bang' Hamiltonians
\cite{bang-bang}. (iii): one may
depart from  a more systematic analysis of
the Lie algebra generated (by commutation)
from a given set of basic Hamiltonians;
If one has access to Hamiltonians $H_1$ and $H_2$
with variable strength parameters $\kappa_1$ and $\kappa_2$, 
subsequent application over short time intervals $dt$ of 
$\kappa_1 H_1$, $\kappa_2 H_2$, $-\kappa_1 H_1$ and $-\kappa_2 H_2$,
leads to the evolution operator ($\hbar =1$)
\begin{eqnarray}
{\rm e}^{i\kappa_2 H_2 dt}{\rm e}^{i\kappa_1 H_1 dt} {\rm e}^{-i\kappa_2
  H_2 dt} 
{\rm e}^{-i\kappa_1 H_1 dt}\nonumber \\
= {\rm e}^{\kappa_1 \kappa_2 [H_1,H_2] dt^2} + O(dt^3),
\label{baker}
\end{eqnarray}
so that effectively the Hamiltonian
$i[H_1,H_2]$ is obtained. 
As expressed by Lloyd \cite{LloydScience}: `By going forward and backing up
a sufficiently small distance a large enough number of times, it is
possible to parallel park in a space only $\varepsilon$ longer than
the length of the car'. If $H_1$ and $H_2$ commute with the commutator
$[H_1,H_2]$, the higher order terms in $dt$ vanish exactly and
one may apply $H_1$ and $H_2$ for arbitrarily large $dt$ and  
'make a round trip in the parking lot and park in one single
operation'. 

The different proposals for quantum computing with
trapped ions can be roughly categorized according to the lines above: 
In their original
proposal\cite{CiracZoller}, 
Cirac and Zoller noted that lasers resonant with sideband
excitation of the trapped ions couple the ground and first
vibrational state conditioned on the internal state of the
irradiated ion, and subsequent irradiation of a second ion 
can couple its internal states conditioned on the vibrational 
state.
We have formulated a proposal  
for two-bit \cite{anders1} and multi-bit \cite{anders2} gates in
the ion trap, which makes use of resonance conditions
to couple certain states of the two-particle system. In our
proposal we apply
bichromatic light which selects certain virtually excited 
intermediate states, and by choosing appropriate parameters
we show  that the desired internal state dynamics of the ions may be
perfectly achieved, even if the vibrational degrees of freedom,
used to couple the ions, are not in their ground state.
Recently, Milburn \cite{milburn} has proposed a realization of a multi-bit 
quantum gate in the ion trap, which also operates when the 
ions are vibrationally excited:
Adjusting the phases of laser fields resonant with side band
transitions, one may couple internal state operators to different
quadrature components,  
{\it e.g.}, position and momentum, $X$ and $P$, of the oscillatory motion. 
In Ref.~\cite{milburn} it is proposed to  use the two Hamiltonians
$H_1=\lambda_1J_zP$ and $H_2=\lambda_2J_z X$, expressed 
in terms of the
collective spin operators $J_\xi=\sum_k j_{\xi k}$ $(\xi=x,y,z)$, where the
sum  is
over the
ions irradiated by the lasers, and where $j_{\xi k}$ 
is the spin operator for the atom $k$, which may be defined by the
Pauli spin matrices $j_{\xi k}=\sigma_{\xi k}/2$ $(\hbar=1)$.
By alternating application of the Hamiltonians $H_1$ and $H_2$ we may obtain
the {\it exact} propagator
\begin{eqnarray}
\label{jzsq}
e^{iH_2 \tau}e^{i H_1 \tau} e^{-i H_2 \tau}
e^{-iH_1 \tau} = e^{-i\lambda_1\lambda_2 J_z^2 \tau^2}
\end{eqnarray}
because the commutator of the oscillator position and momentum is
a number. The interaction contained in $J_z^2$ between the ions has
been established 
via the vibrational degrees of freedom, but after the gate this motion is
returned to the initial state and 
is not in any way entangled with the internal state dynamics. 
Milburn also considers the possibility of coupling different individual
internal state operators 
successively to X and P, so that the commutator term provides the 
product of such operators.

In this paper, we shall demonstrate that our bichromatic excitation
scheme is in fact already a realization of the proposal by 
Milburn, and that 
gate operation more rapid than 
concluded in \cite{anders1} is possible.
We  show that our
bichromatic scheme
implements a propagator of the form ${\rm e}^{-iA(\tau)J_y^2}$ which is
analogous 
to the one obtained by Milburn (\ref{jzsq}).  In Ref.~\cite{anders2} it was
shown that this propagator can be used to prepare maximally
entangled states $\frac{1}{\sqrt2}(|gg...g\ra+{\rm e}^{i\phi} 
|ee...e\ra)$ of any number of ions  ($N$), where the $k$'th letter denotes
the internal state  $e$ or $g$
of the $k$'th ion. These maximally entangled states, which have an interest
in the their own right \cite{ur}, are produced by applying the  unitary
operator ${\rm e}^{i\pi/2 J_y^2}$ to a string of ions initially in the state
$|gg...g\ra$, and they may be produced even without experimental access to
individual ions in the trap. 

In this paper we focus on the preparation of maximally entangled
states. This is both of 
convenience for the theoretical presentation and to emphasize results which
are most easily verified experimentally. However, the procedures described
here  also apply to quantum computation. With two ions illuminated by laser
light the bichromatic scheme produces $\frac{1}{\sqrt2}(|gg\ra-i|ee\ra)$ and
together 
with single qubit rotation this evolution forms a universal set of gates
which may be used to constuct a quantum computer. The {\scshape control-not}
operation \cite{gates} for example may be obtained by
applying single ion operations on each ion before and after the bichromatic
pulse which creates the state $\frac{1}{\sqrt2}(|gg\ra -i |ee\ra)$ from 
$|gg\ra$.

In section \ref{ideal}, we recall our  proposal for a two-qubit gate operation
and we show that it is equivalent to the proposal of Milburn, 
with a harmonic rather than a stroboscopic application of Hamiltonian
coupling terms. 
In experiments it may be difficult to fulfill the requirements for
the analysis of sec.~\ref{ideal} to be precise, and 
in section \ref{nonideal} we address
the fidelity with which certain
entangled states may be engineered when we 
take into account the off-resonant couplings and the finite value of the
Lamb-Dicke parameter. In sec.~\ref{env} we study the influence of the
environment on the system. We analyse the role of spectator vibrational
modes and
energy exchange between the ionic motion and 
thermal surroundings.
A summary of our results and a conclusion are presented in section
\ref{conclusion}.

\section{Gate operation under ideal conditions}
\label{ideal}
Ions in a linear trap interacting with a laser field of frequency $\omega$
may be described by the Hamiltonian
\begin{eqnarray}
  &H =&  H_0+H_{{\rm int}} \nonumber\\ 
  &H_0 =& \nu (a^{\dagger} a+1/2)+\omega_{eg}\sum_i
   \sigma_{zi}/2 \nonumber \\  
  &H_{{\rm int}} =&  \sum_i \frac{\Omega_i}{2}
 (\sigma_{+i}\ {\rm e}^{i(\eta_i(a+a^{\dagger})-\omega t)}+ h.c.),
 \label{hamilton}
\end{eqnarray}
where $\nu$ is the frequency of the vibration, $a^{\dagger}$
and $a$ are the ladder 
operators of the quantized oscillator,  $w_{eg}$ is the
energy difference between the internal states $e$ and $g$,
and  $\Omega_i$ is the resonant Rabi frequency of  the
$i$'th 
ion in the laser field. The exponentials account for the position
dependence of the laserfield, and  the recoil of the ions
upon absorption of a photon. The positions of the ions $x_i$ are replaced by
ladder operators $kx_i=\eta_i(a+a^\dagger)$, where the Lamb-Dicke parameter
$\eta_i$ represents the ratio between 
the ionic excursions within the vibrational ground state wavefunction and
the wavelength  of the
exciting radiation.
In
Eq.~(\ref{hamilton}) we have assumed that the laser is close to a
sideband $\omega\approx\omega_{eg}\pm \nu$ for a single vibrational mode
and  that we may neglect the
contribution from all other 
vibrational modes. We  tune the 
lasers close to the
center-of-mass vibrational mode where all ions participate equally in
the vibration, so that the coupling of the recoil
to the vibration is identical for all ions, {\it i.e.}, $\eta_i=\eta$ for
all $i$. For simplicity we  also assume the same Rabi frequency for
all ions participating in the gate $\Omega_i=\Omega$. In this section we
will  consider an ion trap operating in the
Lamb-Dicke limit, {\it i.e}. the ions are cooled to a regime with vibrational
numbers $n$ ensuring that $(n+1)\eta^2<<1$, so that we may perform the
expansion $e^{i\eta(a+a^\dagger)}\approx 1+i\eta(a+a^\dagger)$.

\begin{figure}[b]
  \noindent
  \epsfig{angle=270,width=\linewidth,file=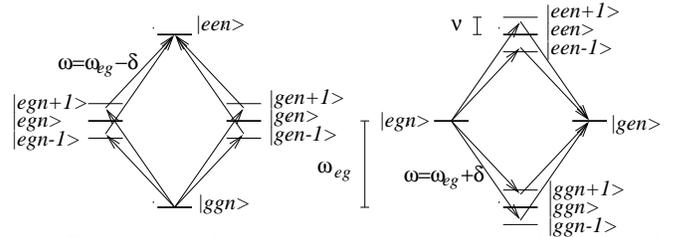}
  \caption{Energy level diagram for two ions with quantized vibrational
    motion illuminated with bichromatic light. The only resonant
    transitions are from $|ggn\ra$ to $|een\ra$ (left)
    and from $|egn\ra$ to $|gen\ra$ (right). Various transition paths
    involving 
    intermediate states with vibrational number $n$ differing by unity are
    identified.}
  \label{detunings}
\end{figure}

\subsection{Weak field coupling}
\label{weak}
In our previous article \cite{anders1} we assumed that two ions in the string
were both illuminated with two lasers of opposite detunings
$\omega-\omega_{eg}=\pm\delta$. With this choice of laser detunings the
only energy conserving transitions are from $|ggn\ra$ to $|een\ra$ and from
$|gen\ra$ to $|egn\ra$, where
$n$ is the 
quantum number for the relevant vibrational mode of the trap, {\it cf.}
Fig.~\ref{detunings}. We considered the weak field
regime 
$\eta\Omega<<\nu-\delta$, where only a negligible population is transfered to
the intermediate levels with vibrational quantum numbers $n\pm1$. In this
regime the effective Rabi frequency $\tilde \Omega$ for the transition from
$|ggn\ra$ to 
$|een\ra$  may be determined in second order
perturbation theory 
\begin{equation}
\tilde \Omega =2{ \sum_m \frac{\langle een|H_{int}|m\ra\langle
                  m|H_{int}|ggn\ra} {E_m-(E_{ggn}+\omega_m)}}
    =- \frac{(\Omega\eta)^2}{\nu-\delta},
\label{omega}
\end{equation}
where we have used the intermediate states  $|m\ra$=$|egn\pm 1\ra$ and $|gen\pm
1\ra$, and where $\omega_m$ is the frequency of the laser  exciting the
intermediate state $|m\ra$. For the transition $|egn\ra$ to $|gen\ra$ we
get the same effective Rabi frequency.

The  remarkable feature in Eq.~(\ref{omega}) is that it contains no
dependence  on the vibrational quantum
number $n$. This is due to interference between the paths shown in
Fig.~\ref{detunings}. If we take a path where an intermediate state
with vibrational quantum number $n+1$ is excited,
we have a factor of $n+1$ appearing in the numerator ($\sqrt{n+1}$ from
raising and  $\sqrt{n+1}$ from lowering the vibrational quantum
number). In
paths involving the vibrational state $n-1$  we obtain
 a factor of $n$. Due to the opposite detunings, the
denominators  have opposite signs and the $n$
dependence disappears when the two terms are subtracted. The coherent
evolution of the internal atomic
 state is thus insensitive to the vibrational quantum
numbers, and it may be observed with ions in any 
superposition or  mixture of 
vibrational states. The coherent evolution may even be seen if the
vibrational quantum number $n$ changes during the gate due to heating
\cite{anders1}. 

\subsection{General field coupling}
\label{strong}
We now consider the  interaction without restricting the parameters to
a regime where no population is transfered to states with different
$n$. For this 
purpose it is convenient to 
change to the interaction picture with respect to $H_0$. In the Lamb-Dicke
limit with lasers detuned by $\pm\delta$ the  Hamiltonian
becomes
\begin{eqnarray}
  \label{interaction}
  H_{\rm int}=&2\Omega J_x&\cos \delta t
  -\sqrt{2}\eta\Omega J_y \nonumber\\
   &&\times  [  x (\cos(\nu-\delta)t+\cos(\nu+\delta)t)\nonumber\\
   && \quad +p
  (\sin(\nu-\delta)t+\sin(\nu+\delta)t) ], 
\end{eqnarray}
where we have introduced the dimensionless position 
and momentum operators, $x=\frac{1}{\sqrt{2}}(a+a^\dagger)$ and
$p=\frac{i}{\sqrt{2}}(a^\dagger-a)$, and the collective spin operators
discussed above Eq.~(\ref{jzsq}). 

Choosing not too strong
laser intensities $\Omega<<\delta$ and tuning close to the sidebands
$\nu-\delta<<\delta$ we may neglect the $J_x$ term  and the terms
oscillating at frequency  $\nu+\delta$ in
Eq.~(\ref{interaction}), and our interaction is a special case of the
Hamiltonian 
\begin{equation}
  \label{xpham}
  H_{\rm int}=f(t) J_y x+g(t) J_y p.
\end{equation}
The exact propagator
for the Hamiltonian~(\ref{xpham}) may be represented by the ansatz
\begin{equation}
  \label{u}
  U(t)={\rm e}^{-iA(t)J_y^2}{\rm e}^{-iF(t)J_y x}{\rm e}^{-iG(t)J_y p},
\end{equation}
and the Schr{\"o}dinger equation  $i\frac{d}{dt}U(t)=H_{\rm int} U(t)$ then
leads to 
the expressions
\begin{eqnarray}
  \label{functions}
  &F(t)=&\int_0^t f(t')dt'\nonumber\\
  &G(t)=&\int_0^t g(t')dt'\nonumber\\
  &A(t)=&-\int_0^t F(t')g(t')dt'.
\end{eqnarray}
With $f(t)=-\sqrt{2}\eta\Omega\cos (\nu-\delta)t$ and
$g(t)=-\sqrt{2}\eta\Omega  \sin
(\nu-\delta)t$ following from (\ref{interaction}) we get
\begin{eqnarray}
\label{fga}
F(t)&=&-{{\sqrt{2}\eta\Omega} \over {\nu-\delta}} \sin{\left((\nu-\delta)t
 \right)}\nonumber \\
G(t)&=&-{{\sqrt{2}\eta\Omega} \over \nu -\delta} {\left[
  1-\cos((\nu-\delta)t)\right]}\nonumber \\
A(t)&=&-{\eta^2\Omega^2 \over \nu -\delta}{\left[t-{1 \over {2
      (\nu-\delta)}} \sin(2(\nu-\delta)t) \right]}.
\end{eqnarray}
In the $xp$ phase space the operator $U$ performs translations
$(x,p)\rightarrow (x+J_yG(t),p-J_yF(t))$ entangled with the internal state
of the ions. 

\begin{figure}[b]
  \begin{center}
    \leavevmode
    \epsfig{angle=270,width=6cm,file=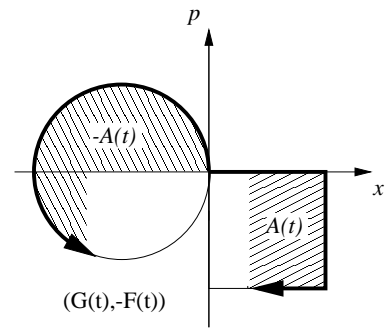}
  \end{center}
  \caption{The paths traversed in phase space and the function $A(t)$ in
    Milburns proposal 
    (rectangular) and in our proposal (circle).}
  \label{fgfig}
\end{figure}

Apart from a change of basis from $J_z$ to $J_y$ the interaction
considered by Milburn\cite{milburn} may also be put in this form, with
$f(t)$ and $g(t)$ alternating between zero and non-vanishing constants.
Within the present formulation,
the trick in Ref.\cite{milburn} is to use functions $f(t)$ and $g(t)$ such
that $F(t)$ and $G(t)$ both vanish after a period $\tau$. At this instant the
vibrational motion is returned to its original state and the propagator
reduces to $U(\tau)={\rm e}^{-iA(\tau)J_y^2}$, {\it i.e.}, we are left with an
internal state evolution which is independent of the external vibrational
state. This decoupling is possible because the effective internal state
transition is completed in the same amount of time for all vibrational
components and because the AC Stark shift of the atomic levels due to the
laser fields are independent of the value of $n$. In the weak field case
these properties are ensured by the interfering coupling amplitude in
Fig.~\ref{detunings}, see detailed discussion in Ref.~\cite{anders1}. In
the general case it follow from the formal structure of
Eq.~(\ref{u}). According  to (\ref{functions}) the acquired factor $A(\tau)$ is
equal to the area swept by the line 
segment between $(G(t),0)$ and $(G(t),-F(t))$, as shown in
Fig.~\ref{fgfig}. If  $(G(t),-F(t))$ forms a closed path,  $A(t)$ is
plus (minus)  the enclosed 
area if the path is traversed in the (counter) clockwise direction. 
In the proposal by Milburn successive constant Hamiltonians proportional to
$x$ and 
$p$ are applied and the area enclosed by $(G(t),-F(t))$ is  rectangular. In
our proposal the area is a circle of radius
$\sqrt{2}J_y \eta\Omega/(\nu-\delta)$, as illustrated in Fig.~\ref{fgfig}.

With the propagator in Eq.~(\ref{u}) we may
calculate the time evolution of the system. Suppose that the ions are
initially in the internal ground state 
and an incoherent mixture of vibrational state as described by the density
matrix $\rho^{tot}=\sum_n P_n |g..gn\ra\langle g..gn|$. The time evolution
of the 
internal state density operator $\rho={\rm Tr}_n(\rho^{tot})$ with any number 
of ions $N$ may be found from
$\rho_{a_1...a_N,b_1..b_N}(t)=\sum_n P_n \langle
g..gn|U^\dagger(t)|b_1..b_N\ra\langle 
a_1..a_N|U(t)|g..gn\ra$ ($a_j,b_j=e$ or $g$), where we have used $\sum_n
|n\ra\langle n|=1$ to remove 
one of the summations over vibrational states. Here we list  the  density
matrix  elements for the case of two ions $N=2$:
\begin{eqnarray}
  \label{density}
  \rho_{gg,gg}&=&\sum_n P_n {\text {\huge [}} \frac{3}{8} +
      \frac{1}{2} {\rm
      e}^{-\frac{F(t)^2+G(t)^2}{4}}\nonumber\\&&\times L_n{\left(
      \frac{F(t)^2+G(t)^2}{2}\right)} 
      \cos{\left( A(t)+\frac{1}{2}F(t)G(t)\right)}\nonumber\\
      && + \frac{1}{8} {\rm e}^{-(F(t)^2+G(t)^2)}
        L_n{\left( 2(F(t)^2+G(t)^2)\right)}  {\text {\huge ]}} \nonumber\\
  \rho_{ee,ee}&=&\sum_n P_n {\text {\huge [}} \frac{3}{8} -
      \frac{1}{2} {\rm
      e}^{-\frac{F(t)^2+G(t)^2}{4}}\nonumber\\&&\times
    L_n{\left(\frac{F(t)^2+G(t)^2}{2}\right)}
      \cos{\left(A(t)+\frac{1}{2}F(t)G(t)\right)}\nonumber\\
      && + \frac{1}{8} {\rm e}^{-(F(t)^2+G(t)^2)}
        L_n(2(F(t)^2+G(t)^2))  {\text {\huge ]}} \nonumber\\
  \rho_{gg,ee}&=&\sum_n P_n {\text {\huge [}} \frac{1}{8}(1- {\rm
    e}^{-(F(t)^2+G(t)^2)} \nonumber\\
    &&\qquad \times L_n(2(F(t)^2+G(t)^2)))\nonumber\\
    &&-\frac{i}{2}
      {\rm e}^{-\frac{F(t)^2+G(t)^2}{4}}
      L_n{\left(\frac{F(t)^2+G(t)^2}{2}\right)} \nonumber \\ 
    &&\qquad \times \sin{\left(A(t)+\frac{1}{2}F(t)G(t)\right)} {\text
      {\huge ]}}, 
\end{eqnarray}
where $L_n$ is the $n$'th order Laguerre polynomium.

\begin{figure}
\begin{center}
 \begin{minipage}{4.4cm}
  \epsfig{file=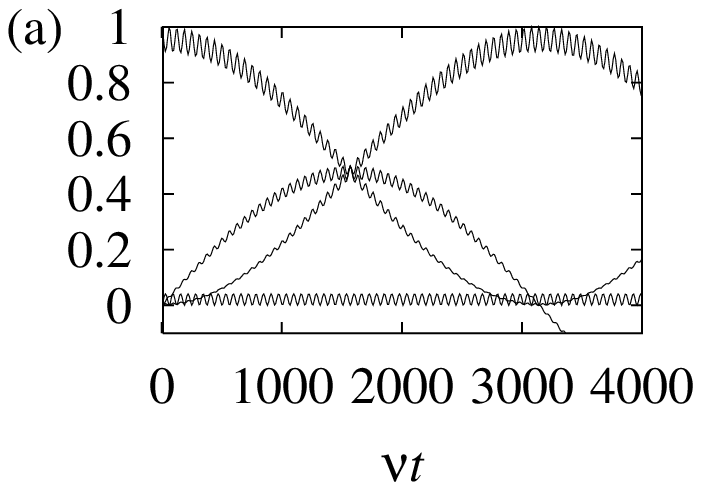,angle=0,width=4.3cm}
 \end{minipage}
\hspace{-0.38cm}
 \begin{minipage}{4.4cm}
  \epsfig{file=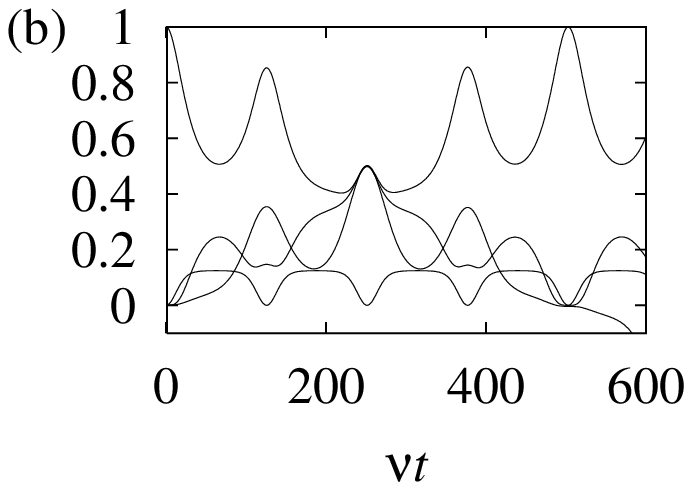,angle=0,width=4.3cm}
 \end{minipage}
\end{center} 
  \caption{Time evolution of density matrix elements for two ions
    calculated from 
    Eq~(\ref{density}). (a) Weak field regime (b) Fast gate. The first
    curve (counting
    from above at $\nu t\approx 1000$ in (a) and $\nu t\approx130$ in (b))
    represents $\rho_{gg,gg}$, the second is the
    imaginary part of $\rho_{gg,ee}$, the third is $\rho_{ee,ee}$, and the
    last curve is the real part of $\rho_{gg,ee}$. The ions are initially
    in the internal state 
    $|gg\ra$ and a thermal vibrational state with an average of 2
    vibrational quanta. 
    In (a) the physical
    parameters are $\delta=0.9\nu$, $\eta=0.1$, and $\Omega=0.1\nu$. In (b)
    the physical parameters are $\delta=0.95\nu$, $\eta=0.1$, and
    $\Omega=0.177\nu$. The parameters in (b) are chosen such that a
    maximally entangled
    state $\frac{1}{\sqrt2}(|gg\ra-i|ee\ra)$ is formed at the
    time $\nu t\approx250$, where the circular path in Fig.~\ref{fgfig} has
    been traversed twice.} 
  \label{twogates}
\end{figure}

These expressions can be evaluated in different regimes.
In the weak field regime,
$\eta\Omega<<\nu-\delta$, 
the $xp$ phase space trajectory is a very small circle, which is traversed
several 
times.  $F(t)$ and $G(t)$ are negligible for all times, and
${\rm 
  e}^{-iF(t)J_y x}{\rm e}^{-iG(t)J_y p}$ is approximately unity, such that
we have an internal state preparation which is disentangled from the
vibrational motion throughout the gate. Since
$A(t)\approx-\eta^2\Omega^2t/(\nu-\delta)$ if 
$(\nu-\delta)t>>1$ the time evolution corresponds to the one
obtained from an effective Hamiltonian $H=\tilde{\Omega}J_y^2$, and
Eq.~(\ref{density}) describes simple Rabi oscillations between the states
$|gg\ra$ and $|ee\ra$. This is demonstrated in Fig.~\ref{twogates}~(a)
which shows the time evolution described by Eq.~(\ref{density}).
The curves show
sinusoidal Rabi oscillation from $|gg\ra$ to $|ee\ra$ superimposed by small
oscillations due to the weak entanglement with the vibrational motion.

\noindent

Outside the weak field regime the internal
state is strongly entangled with the
vibrational motion in the course of the gate. For successful gate
operation   
we have to ensure that we 
return to the 
initial vibrational state at the end of the gate by choosing
parameters such that $G(\tau)=F(\tau)=0$,
corresponding to $(\nu-\delta)\tau=K2\pi$,
where $K$ is an integer. A maximally entangled state 
is created if we adjust our parameters so that $A(\tau)=-\pi/2$. This is
achieved if the parameters are chosen in accordance with 
\begin{equation}
 {\eta\Omega \over \nu-\delta}={1 \over 2\sqrt{K}},\quad K=1,2,3,...\ .
 \label{krav}
\end{equation}
By choosing a low value of $K$ such that an entangled state is created
after a few rounds in phase space we may perform a faster gate than
considered in the weak field case. See
Fig.~\ref{twogates}~(b), where we have used $K=2$, and where a 
maximally entangled state $\frac{1}{\sqrt2}(|gg\ra-i|ee\ra)$ is created 
at the time $\nu t \approx 250$.

By combining the requirement $(\ref{krav})$ with the condition
$(\nu-\delta) 
\tau = K2 \pi$ we may express the time for the state preparation as 
\begin{equation}
 \tau=\frac{\pi}{\eta\Omega}\sqrt{K}. 
\label{tid}
\end{equation}
In order to avoid off-resonant excitations of the ions we must require
$\frac{\Omega^2}{\nu^2}<<1$  and  $\eta^2$ must be much less than
unity 
to fulfill the Lamb-Dicke approximation (see subsec.  \ref{sec:carrier} and
\ref{lambdicke}). For a given 
trap and/or laser intensity Eq.~(\ref{tid}) sets a bound on the speed
of the gate. In tabel \ref{tabel:tid} we give some numerical examples
for the time of the gate for some typical experimental parameters.
The {\sc control-not} operation which plays a central role in quantum
computation \cite{gates} may be created by a combination of single
particle rotations and a bichromatic pulse with the duration described 
by Eq.~(\ref{tid}). The single particle operations may be performed much
faster than the two qubit gates, so the time required to perform a
{\sc control-not} operation is also given by (\ref{tid}).

\newcommand{\PreserveBackslash}[1]{\let\temp=\\#1\let\\=\temp}
\let\PBS=\PreserveBackslash
\PBS
\noindent
\setlength{\extrarowheight}{9 pt}
\begin{table}[bt]
\caption{The time required to create the maximally entangled state
  $\frac{1}{\sqrt{2}} (|gg...g\ra-i|ee..e\ra)$ with a Lamb-Dicke
    parameter $\eta=0.1$ for various trapping frequencies ($\nu$) and
    laser intensities ($\Omega$). The table shows the gate time if the
    entangled state is prepared after a single round in phase
    space. If the gate operation is accomplished after $K$ rounds in
    phase space the time should be multiplied by $\sqrt{K}$.}  
\label{tabel:tid}
\vspace{0.2cm}
\begin{tabular*}{\linewidth}{m{1.5cm}
    >{\centering}m{2.5cm}
    >{\centering}m{2.2cm}  
    >{\PBS\centering}m{2.2cm}}
\hline \hline
~ $\frac{\Omega}{\nu}$ \qquad &
  $\frac{\nu}{2\pi}=$500 KHz& 1 MHz&  
  10 MHz \\
0.05 & 200 $\mu$s & 100$\mu$s& 10
  $\mu$s \\ 
0.10 & 100 $\mu$s & 50 $\mu$s& 5 $\mu$s\\ 
0.20 &  50 $\mu$s & 25 $\mu$s& 2.5 $\mu$s\vspace{0.2cm}\\ 
\hline \hline
\end{tabular*}
\end{table}

\section{Non ideal conditions}
\label{nonideal}
In the previous section we used the Lamb-Dicke and the rotating
wave approximations to arrive at an exactly solvable model. In this section 
we perform a more detailed analysis of the validity of the approximations and
we estimate the effect of deviations from the ideal situation in an actual
experiment.  The 
general procedure in the section, is to change to the interaction picture
with respect to the simple Hamiltonian (\ref{xpham}) using the exact
propagator in 
Eq.~(\ref{u}), and to treat the small deviations from the ideal situation
by perturbation theory. The
figure of  merit for the performance of the gate is taken to
be the fidelity $F$ of creation of the maximally  entangled $N$-particle
state $|\Psi_{max}\ra=1/\sqrt{2}(|gg..g\ra-i|ee..e\ra)$,
which in the ideal case is created
at the time when $A(\tau)=-\pi/2$, if the ions are initially in the
$|gg..g\ra$ state \cite{anders2}, 
{\it i. e.}, 
\begin{equation}
 F=\langle\Psi_{max}|\rho_{int}(\tau)|\Psi_{max}\ra. 
\label{fidelity.def}
\end{equation}

\subsection{Direct coupling}
\label{sec:carrier}
Going from Eq.~(\ref{interaction}) to Eq.~({\ref{xpham}) we neglected a term
$H_{d}=2\Omega J_x \cos(\delta t)$. This term describes direct off resonant
coupling of $g$ and $e$
without changes in the vibrational motion. For high laser
power this term has a detrimental effect on the fidelity, which we
calculate below.

Changing to the interaction picture, we may find the propagator $U_I(t)$
from the Dyson series
\begin{eqnarray}
 \label{dyson}
 U_I(t)&=&1-i\int_0^t dt' H_{d,I}(t')\nonumber \\
  &&\quad-\int_0^t\int_0^{t'} dt' dt''
 H_{d,I}(t')H_{d,I} (t'')+\dots ~,
\end{eqnarray}
where the interaction Hamiltonian is given by $H_{d,I}(t)=U^\dagger
(t)H_d(t)U(t)$. Since $H_d(t)$ is oscillating at a much higher frequency
than the propagator $U(t)$,  we may treat $U(t)$ as a
constant during the integration and we obtain
\begin{eqnarray}
 \label{ucarrier}
 U_I(t)&=&1-i\frac{2\Omega}{\delta}\sin(\delta
 t)U^\dagger(t)J_xU(t)\nonumber\\
 &&\quad -\frac{\Omega^2}{\delta^2} (1-\cos(2\delta t))
 U^\dagger(t)J_x^2U(t) + \dots ~.
\end{eqnarray}
Near the endpoint, $U(t)\approx {\rm e}^{i(\pi/2) J_y^2}$ and we obtain the
fidelity
\begin{equation}
 \label{fidelc}
 F\approx 1- \frac{N\Omega^2}{2\delta^2}(1-\cos(2\delta \tau)),  
\end{equation}
where $N$ is the number of ions participating in the gate.  
We plot in Fig.~\ref{carrierfig} the product of the fidelity due to the carrier
(\ref{fidelc}) and the population of the  
EPR-state $\frac{1}{\sqrt2}(|gg\ra-i|ee\ra)$ expected from  the time
evolution in 
Eq.~(\ref{density}). The result agrees well with the result of a 
numerical integration of 
the Schr{\"o}dinger equation with the Hamiltonian (\ref{interaction}).

\begin{figure}[htbp]
  \begin{center}
    \leavevmode
    \epsfig{file=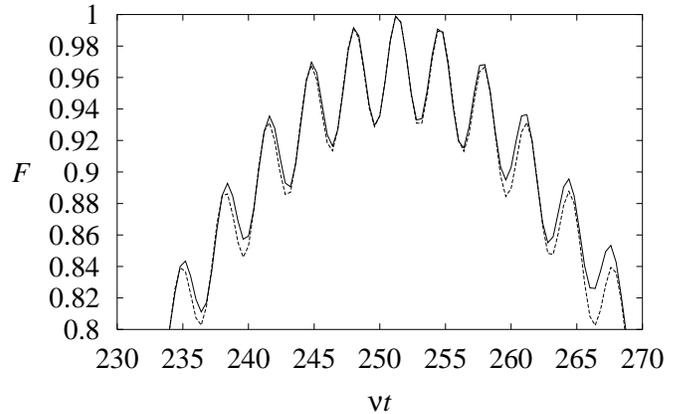,angle=0,width=\linewidth}
  \end{center}
  \caption{Population of the EPR-state $\frac{1}{\sqrt2}(|gg\ra-i|ee\ra)$
    near the 
    optimum. The full line is obtained by a numerical integration of the
    Hamiltonian (\ref{interaction}) and the dashed line is the product of the
    expression in (\ref{fidelc}) and the expression for the fidelity
    obtained from Eq. (\ref{density}). The parameters are the same as in
    Fig.~\ref{twogates}~(b).}
  \label{carrierfig}
\end{figure}

If the duration of the laser pulses can be controlled very accurately in 
the experiment, so that one fulfills both (\ref{krav}) and $2\delta \tau=2
K'\pi$  the effect of the direct coupling vanishes. If one cannot perform such
an accurate control, the net effect of the direct coupling is to reduce the
average fidelity by $\frac{N\Omega^2}{2 \delta^2}$ (=0.03 for the
parameters of Fig.~\ref{carrierfig}).

\subsection{Lamb-Dicke approximation}
\label{lambdicke}
In section \ref{ideal} we used the Lamb-Dicke approximation ${\rm e}^{i\eta
  (a+a^\dagger)}\approx 1+ i\eta (a+a^\dagger)$ to simplify our
calculations. Now we investigate the validity of this approximation. 

In the weak field case, we can use the exact matrix elements $\langle
n|{\rm e}^{i\eta 
  (a+a^\dagger)}|n+1\ra=i\eta\frac{{\rm
    e}^{-\eta^2/2}}{\sqrt{n+1}}L^1_n(\eta^2)$, to obtain the effective
Rabi frequency between $|ggn\ra$ and $|een\ra$
\begin{eqnarray}
  \label{omegan}
\tilde{\Omega}_n&=&\tilde{\Omega}
 {\rm e}^{-\eta^2}
 {\left[ 
  {{\left(L_n^1(\eta^2) \right)^2} \over {n+1}}- {{\left(L_{n-1}^1(\eta^2)
  \right)^2}  \over {n}} \right]}\nonumber \\
&\approx&\tilde{\Omega}{\left[ 1-\eta^2 (2n+1)+\eta^4 
 {\left (\frac{5}{4}n^2+\frac{5}{4}n+\frac{1}{2} \right)}\right]},
\end{eqnarray}
where $\tilde{\Omega}$ is given by Eq.~(\ref{omega}), and where $L_n^1$ are
the generalized Laguerre polynomials
\begin{equation}
 \label{laguerre}
 L_n^\alpha(x)=\sum_{m=0}^{n}(-1)^m 
 {\left(  
 \begin{array}{cc}
  {n+\alpha} 
   \\ {n-m}
 \end{array} \right)}
 \frac{x^m}{m!}. 
\end{equation} 
The effective Rabi frequency is no longer independent of the vibrational
quantum number $n$, and the internal state becomes entangled with the
vibrational motion, resulting in a non-ideal performance of the gate
\cite{zagury}. 

To illustrate the effect of deviations from the Lamb-Dicke approximation,
we consider again the production of an EPR-state
$\frac{1}{\sqrt2}(|gg\ra-i|ee\ra)$.  With an
$n$-dependent coupling strength the fidelity is
\begin{equation}
 F=\frac{1}{2}+\frac{1}{2}\sum_{n=0}^{\infty}P_n \sin(\tilde{\Omega}_n t),  
 \label{fidelld}
\end{equation}
where $P_n$ is the initial population of the vibrational state
$n$. We show in Fig.~\ref{fidelldfig} the evolution of the
fidelity predicted by Eq.~(\ref{fidelld}) and obtained by a direct
integration of the full Hamiltonian in Eq.~(\ref{hamilton}). 
Due to the deviation from the Lamb-Dicke approximation the effective Rabi
frequency is reduced, {\it cf.},  Eq.~(\ref{omegan}),  and the
optimal gate performance is achieved with a duration that is longer
than $\pi/(2\tilde{\Omega})$. The spreading of the values of
$\tilde{\Omega}_n$, causes entanglement with
the vibrational motion which reduces the fidelity. With the
parameters in Fig.~\ref{fidelldfig} the maximally obtainable fidelity is
0.92 obtained after a pulse of duration $\tau\approx1.9/\tilde{\Omega}$. 
\begin{figure}[tbp]
    \begin{center}
      \leavevmode
    \epsfig{angle=270,width=\linewidth,file=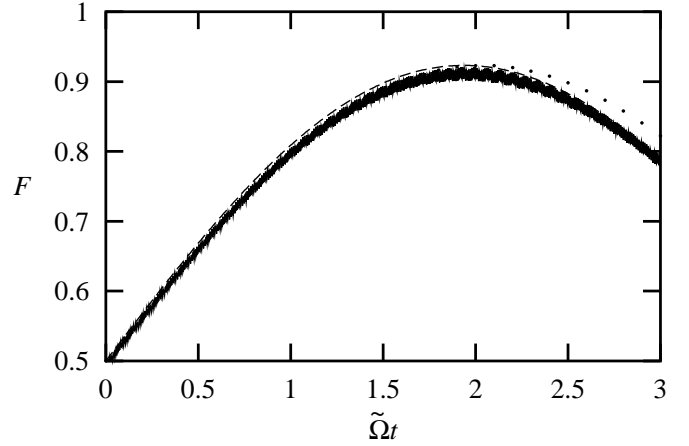}
    \end{center}
    \caption[]{Evolution of the population of the EPR-state
      $\frac{1}{\sqrt2}(|gg\ra-i|ee\ra)$ 
      for a vibrational thermal state with an average of 5 vibrational
      quanta and 
      $\eta=0.20$. The dotted line is the prediction from
      Eq.~(\ref{fidelld}) and the solid line is the result of a numerical
      integration of the Hamiltonian~(\ref{hamilton}) with parameters
      $\Omega=0.02 \nu$ and $\delta=0.9 \nu$. The discrepancy between
       the two curves at $\tilde{\Omega}t\gtrsim2$ is due to additional
       off-resonant couplings which may be taken into account by
       multiplying  the coupling strength by $\frac{2\nu}{\nu+\delta}$
       (dashed curve) \cite{anders1,anders2}.}
    \label{fidelldfig}
\end{figure}

With more than two ions, the time evolution of the system may be obtained by
expanding the  initial state $|gg...g\ra$ on eigenstates of the $J_y$
operator: 
\begin{eqnarray}
|gg...g\ra= 
 \frac{(-i)^N}{2^{N/2}} \sum_{k=0}^N (-1)^k 
 \sqrt{  {\left(   
 \begin{array}{cc}
  {N} 
   \\ {k}
 \end{array} \right)}}
|M_y=N/2-k\ra.
\label{expansion} 
\end{eqnarray}
In the $J_y$ basis the propagator (\ref{u}) is diagonal
and in the weak field regime ($F(t)$, $G(t)\approx 0$) with $n$ dependent
coupling strengths we get the  fidelity
\begin{equation} 
 \label{fidelldmany}
 F={\sum_{n=0}^\infty} P_n
 {\left| \frac{1}{2^N} \sum_{k=0}^N  {\left(   
 \begin{array}{cc}
  {N} 
   \\ {k}
 \end{array} \right)} {\rm e}^{i(N/2-k)^2 (\pi/2-\tilde{\Omega}_n t)}
\right|}^2.   
\end{equation}
In the limit of many ions $(N>>1)$ and near the optimum ($\tilde{\Omega}_n t
\approx \pi /2$) we may approximate this expression by assuming that $k$ is
a continuous variable and by replacing the binominal coefficient by
a Gaussian distribution with the same width. In this limit the fidelity
becomes 
\begin{equation}
  \label{fidelldmanygauss}
  F=\sum_{n=0}^\infty P_n {{1}\over
    {\sqrt{1+\frac{N(N-1) (\pi/2-\tilde{\Omega}_n 
          t)^2}{4}}}}.
\end{equation}
Expanding this  expression to lowest order in $\eta$ and adjusting the
pulse duration to take into account the reduction in the coupling strength
we find to lowest order in $\eta$ 
\begin{equation}
  \label{fidelldlowest}
  F=1-\frac{\pi^2N(N-1)}{8} \eta^4 {\rm Var}(n)   
\end{equation}
at the optimum time
\begin{equation}
  \tau_{opt}={\pi \over 2 \tilde{\Omega}} (1+\eta^2(2 \bar{n}+1)),
  \label{ldtopt}
\end{equation}
where $\bar{n}$ and Var$(n)$ are the mean and variance of the vibrational
quantum number.  

In Eq. (\ref{fidelldmanygauss}) and (\ref{fidelldlowest}) we have
replaced a quantity $N^2$ following from the Gaussian approximation to
(\ref{fidelldmany}) by 
$N(N-1)$. With this substitution (\ref{fidelldmanygauss}) and 
(\ref{fidelldlowest}) describe the fidelity well for all values of
$N$. With the parameters of Fig.~\ref{fidelldfig}, Eq.~(\ref{fidelldlowest})
yields $F=0.88$  which is in good
agreement with 
the numerical result in
the figure. 
 
The equations (\ref{omegan}-\ref{fidelldlowest}) were derived for weak
fields, but they also provide an accurate description of  the system
outside this regime.  To show this we note, that with bichromatic light,
$H_{\rm int}$ in Eq.~(\ref{hamilton}) may be written as   
\begin{eqnarray}
  \label{hnonld}
  H_{\rm int}&=&2\Omega\cos(\delta t){\Big[} J_x \cos{\left( \eta\sqrt{2}(x
      \cos(\nu t)+p\sin(\nu t))\right)}\nonumber \\
      &&\quad -J_y \sin{\left( \eta\sqrt{2}(x
      \cos(\nu t)+p\sin(\nu t))\right)}{\Big]}
\end{eqnarray}
in the
interaction picture with respect to $H_0$. An expansion of the trigonometric
functions in this Hamiltonian
leads to Eq.~(\ref{interaction}) which formed the
basis of the discussion in section \ref{ideal}. The term proportional to
$J_x$ is  suppressed because it is   far off
resonance.  The lowest
order contribution of this term was treated in the
previous section, and we shall now consider corrections to the $J_y$ term
which may have
significant effects. In the interaction
picture with respect to the lowest order Hamiltonian~(\ref{xpham}),  $x$
and $p$ are changed into 
$x+J_y G(t)$ and $p-J_y F(t)$ and to lowest non-vanishing order in $\eta$ the
interaction picture Hamiltonian is 
\begin{eqnarray}
  \label{Heta3}
  H_3&=&\eta^3 J_y \frac{\sqrt{2}\Omega}{12}\Big[ \cos((\nu-\delta)t)h_1(x,p)
  \nonumber \\
  && \qquad \qquad \qquad +\sin((\nu-\delta)t) h_2(x,p)\Big],
\end{eqnarray}
where
\begin{eqnarray}
  \label{h1h2}
  h_1(x,p)&=&3x^3+xp^2+pxp+p^2x \nonumber \\
  h_2(x,p)&=&3p^3+px^2+xpx+x^2p,
\end{eqnarray}
and where we have used that $F(t)$ and $G(t)$ are proportional to $\eta$. 
To calculate the effect of the Hamiltonian in (\ref{Heta3}) we note that the
propagator 
\begin{eqnarray}
  \label{u3}
  U_{3,{\rm int}}(t)={\rm
    e}^{{\left[-i\frac{\sin((\nu-\delta)t)} {\nu-\delta}h_1(x,p) \right]}} 
    {\rm e}^{{\left[-i\frac{1-\cos((\nu-\delta)t)} {\nu-\delta}h_2(x,p)
    \right]} }
\end{eqnarray}
is  consistent with the Hamiltonian (\ref{Heta3}) until order $\eta^5$,
{\it i.e.} $i\frac{dU_{3,{\rm int}}(t)}{dt}=(H_3+O(\eta^6))U_{3,{\rm
    int}}$. (But the full Hamiltonian contains terms of order $\eta^4$ and
$\eta^5$ which are not taken into account in $U_{3,{\rm int}}$. These terms
are included below). We are
interested  in
the propagator at times
$\tau=K2\pi/(\nu-\delta)$ where the vibrational motion is returned to the
initial state. At these instants the exponents in Eq.~(\ref{u3}) vanish
and the propagator reduces to $U_3(\tau)=1$ such that it has no influence on
the internal state preparation.

 Expanding the Hamiltonian to order $\eta^6$ we 
obtain the propagator to the same order in $\eta$ in the interaction picture
with respect to $H_0$ in 
(\ref{hamilton}) 
\begin{eqnarray}
\label{u6}
  U_6(\tau)&=&{\rm e}^{-i\tilde{\Omega}\tau J_y^2{\left[ 1-\eta^2
      (2n+1)+\eta^4  
 {\left (\frac{5}{4}n^2+\frac{5}{4}n+\frac{1}{2} \right)}\right]}}
 \nonumber\\&&\quad\times {\rm e}^{i\eta^5
   J_y^3\frac{\sqrt{8}\Omega^3}{(\nu-\delta)^2}x \tau}
  {\rm e}^{-i \eta^6 J_y^4 \frac{5\Omega^4}{2(\nu-\delta)^3} \tau}
\end{eqnarray}
valid at times $\tau=K2\pi/(\nu-\delta)$. The first exponential  provides the
time evolution with the modified effective Rabi frequency in
Eq.~(\ref{omegan}). If we evaluate the propagator (\ref{u6}) in
the weak field regime, the
last two 
exponentials  both vanish in the limit of large K 
when  the requirement (\ref{krav}) is inserted, and the 
time evolution in (\ref{u6}) is consistent with
Eq.~(\ref{omegan}-\ref{fidelldlowest}). 
The last two exponentials are also of minor importance for a different
reason: In  Eq.~(\ref{omegan})  $\eta^2$ appears
in the 
combination $\eta^2 n$, whereas it appears as $\eta^2$ in the last two
exponentials of (\ref{u6}) when the condition (\ref{krav}) is inserted. In
situations 
where  deviations from the Lamb-Dicke
approximation are important $\eta^2 n \sim 1$, the deviation is typically
caused by a high value of $n$ rather than a large value of $\eta$
($\eta^2<<1$). In this case one may neglect the last two exponentials and
the effect of the non-Lamb-Dicke terms are the same as in the case of
weak fields as described by Eqs.~(\ref{omegan}-\ref{fidelldlowest}).  To
achieve the  optimum operation of the gate
with the parameters of Fig.~\ref{fidelldfig} we have to ensure
$\tilde{\Omega}\tau\approx 1.9$ and there is a small correction to the
condition in Eq.~(\ref{krav}). 

\section{External disturbances}
\label{env}
So far we have considered a system described by the Hamiltonian
(\ref{hamilton}), 
where only the center of mass motion is present in the ion trap and where
the coupling of this mode to the surroundings is neglible. In this section
we shall remove these two assumptions and consider the decrease in fidelity
due to the presence of other modes in the trap and due to 
heating of the center of mass vibrational motion.

\subsection{Spectator vibrational modes}
\label{spectator}
With $N$ ions in the trap, the motional state is described by
$3N$ non degenerate vibrational modes. With a proper laser geometry or if
the transverse potential is much steeper than the longitudinal potential, 
the coupling of the laser to transverse modes will be neglible and the only
contribution is from the $N$ longitudinal modes. With $N$ vibrational modes
the ion trap may be
described by the Hamiltonian
\begin{eqnarray}
  &H =&  H_0+H_{{\rm int}} \nonumber\\ 
  &H_0 =&\sum_{l=1}^N \nu_l (a_l^{\dagger} a_l+1/2)+\omega_{eg}\sum_i
   \sigma_{zi}/2 \nonumber \\  
  &H_{{\rm int}} =&  \sum_{i=1}^N \frac{\Omega_i}{2}
 (\sigma_{+i}\ {\rm e}^{i(\sum_{l=1}^N \eta_{i,l}(a_l+a_l^{\dagger})-\omega
   t)}+ h.c.), 
 \label{hammany}
\end{eqnarray}
where $\nu_l$ and $a_l^{\dagger}$ and $a_l$ are the frequency and ladder
operators of the $l$'th mode. The excursion of the $i$'th ion in the $l'th$
mode is described by the Lamb-Dicke parameter $\eta_{i,l}$ which may be
represented as $\eta_{i,l}=\eta \frac{\sqrt{N}b_i^l}
{\sqrt{\nu_l/\nu}}$, where $\eta$ and $\nu$ refer to the center of mass
mode as in the previous sections, and where $b_i^l$ obeys the orthogonality
conditions $\sum_{i=1}^N b_i^l b_i^{l'}=\delta_{l,l'}$ and  $\sum_{l=1}^N
b_i^l b_{i'}^{l}=\delta_{i,i'}$ \cite{james}. 

The center of mass mode ($l=1$), which is used to create the entangled
states of the ions,  has $b_i^1=1/\sqrt{N}$ for all ions 
and is well isolated from the remaning $N-1$ vibrational modes $\nu_{l>1}\geq
\sqrt{3} 
\nu$, so that we could neglect the contribution from the other modes in the
previous sections. In this section we shall extimate the effect of the
presence of the 
spectator modes. They have both a direct effect, due
to the off resonant coupling to the other modes, and an indirect
'Debye-Waller'  effect \cite{nist} because the coupling strength of the center
of mass mode is reduced
due to the oscilations in the spectator modes. Below we shall 
calculate the direct and indirect effects separately. 

The lowest order contribution of the direct coupling to the spectator modes
may be found by expanding the exponentials as in
Eq.~(\ref{interaction}). 
\begin{eqnarray}
  \label{xpmany}
  H_{\rm int}=&2\Omega J_x&\cos \delta t+\sum_{l=1}^N \Theta_l [x_l
  f_l(t)+p_l g_l(t)],
\end{eqnarray}
where $f_l(t)=-\sqrt{2}\eta\Omega \sqrt{\nu/\nu_l}[\cos(\nu_l-\delta)t
+\cos(\nu_l+\delta)t]$ and $g_l(t)=-\sqrt{2}\eta\Omega
\sqrt{\nu/\nu_l}[\sin(\nu_l-\delta)t 
+\sin(\nu_l+\delta)t]$, and where the internal and external state operators
are defined by 
$\Theta_l= \sum_{i=0}^N b_i^l j_{y,i}$  and
$x_l=\frac{1}{\sqrt{2}}(a_l+a_l^\dagger)$ and 
$p_l=\frac{i}{\sqrt{2}}(a_l^\dagger-a_l)$.
Since the ladder operators for different modes
commute, we may find the propagator for this Hamiltonian using the steps that
lead  to Eq.~(\ref{u})
\begin{equation}
 U(t)=\prod_{l=1}^N U_l(t),
 \label{uprod}
\end{equation}
where 
\begin{equation}
  U_l(t)={\rm e}^{-iA_l(t)\Theta_l^2}{\rm e}^{-iF_l(t)\Theta_l x_l}{\rm
    e}^{-iG_l(t)\Theta_l p_l} 
\label{ul}
\end{equation}
with the functions
$F_l$, $G_l$ and $A_l$ defined analogously to Eq.~(\ref{functions}). Note,
that this is an exact solution of the Hamiltonian (\ref{xpmany}) without
the $J_x$ term, so that to
lowest order in the Lamb-Dicke parameter it includes all effects of the
coupling to the  other modes. 

From the definition of $\Theta_l$ it is seen that $\Theta_1=J_y$ and the
propagator $U_1$ reduces to Eq.~(\ref{u}) in the rotating wave
approximation. The  other $N-1$ propagators
in (\ref{uprod}) cause a reduction of the fidelity due to the excursion
into the $x_l p_l$ phase space of these modes. Expanding the exponentials,
using 
$\langle gg...g|\Theta_l \Theta_{l'}|gg...g\ra=\delta_{l,l'}N/4$ and
$\delta\approx\nu$, and    
averaging over time we find
\begin{equation}
 \label{fidelother}
F=1-\eta^2 N \frac{\Omega^2}{\nu^2}\sum_{l=2}^N \frac{\nu}{\nu_l}
(2\bar{n}_l+1)\frac{\nu_l^2/\nu^2+1}{(\nu_l^2/\nu^2-1)^2}, 
\end{equation}
where $\bar{n}_l$ is the mean vibrational excitation of the $l'th$ mode. 

In addition  to the direct coupling to the spectator vibrational mode, the
fidelity is also reduced because the coupling strength is dependent on the
vibration of the other modes. Unlike the direct coupling discussed above,
this effect is not suppressed by the other modes being far off-resonant,
and it  may have
an effect comparable to the direct coupling.  

Due to the vibration of the ions the coupling of the $i$'th ion to the sideband
is reduced from $i\eta\sqrt{n+1}$ to $\langle n_1 n_2 ...n_N|{\rm
  e}^{i\sum_{l=1}^N \eta_{i,l}(a_l+a_l^{\dagger})}|  n_1+1 n_2
...n_N\rangle\approx i\eta\sqrt{n+1} (1-\sum_{l=1}^N
\eta_{i,l}^2(n_l+1/2))$. With this reduced coupling strength the effective
propagator at times $\tau=K2\pi/(\nu-\delta)$ may be described by 
\begin{equation}
 U(\tau)={\rm e}^{-iA(\tau)\Lambda^2},
 \label{udebye}
\end{equation}
where $\Lambda=\sum_{i=1}^N j_{y,i}(1-\sum_{l=1}^N\eta_{i,l}^2
(n_l+1/2))$. In the Cirac-Zoller scheme \cite{CiracZoller},
the $n$-dependent AC Stark shifts caused by coupling to other vibrational
modes lead to decoherence, unless these modes are cooled to the ground
state. In our bichromatic scheme, these internal state level shifts depend
much less on the vibrational excitation.
By expanding (\ref{udebye}) around the optimum
$A(t)\approx\pi/2$ we calculate the lowest order
reduction in the fidelity 
\begin{eqnarray}
  F&=&1-\frac{\pi^2N(N-1)}{8} \eta^4 \sum_{l=1}^N \frac{{\rm Var}(n_l)}
  {(\nu_l/\nu)^2} \nonumber \\ 
  &&\quad-\frac{\pi^2(N-2)}{16} \eta^4\sum_{i,l,l'=1}^N
  \frac{(b_i^l)^2(b_i^{l'})^2-1/N^2}{\nu_l\nu_{l'}/\nu^2} \overline{n_l
    n_{l'}}. 
\label{fideldebye}
\end{eqnarray}

The expressions in Eqs. (\ref{fidelother}) and (\ref{fideldebye}) may be
simplified if the vibrational motion is in a thermal equilibrium at a given
temperature. In a thermal state
Var$(n_l)=\bar{n}_l^2+\bar{n}_l$,
$\overline{n_ln_{l'}}=\bar{n}_l\bar{n}_{l'}$ for $l\neq l'$, and $\bar{n}_l
\le \bar{n}_{1}\nu/\nu_l$, 
and using these expressions we find the lower estimate for the fidelity
\begin{equation}
 F\geq 1-\eta^2 N \frac{\Omega^2}{\nu^2} (\bar{n}_1\sigma_1(N)+\sigma_2(N))
 \label{fidelotherterm}
\end{equation} 
for the direct coupling (\ref{fidelother}) and 
\begin{eqnarray}
  F&\geq&1-\frac{\pi^2N(N-1)}{8} \eta^4 (\bar{n}_1^2 \sigma_3(N)+ \bar{n}_1
  \sigma_4(N)) \nonumber \\  
  &&\quad-\frac{\pi^2(N-2)}{16} \eta^4(\bar{n}_1^2 \sigma_5(N)+ \bar{n}_1
  \sigma_6(N)) 
\label{fideldebyeterm}
\end{eqnarray}
 for the Debye-waller coupling (\ref{fideldebye}), where the sums
 $\sigma_1...\sigma_6$ may be derived from  Eqs. (\ref{fidelother}) and
 (\ref{fideldebye}). For example $\sigma_3(N)=\sum_{l=1}^N
 \frac{\nu^4}{\nu_l^4}$. With the mode functions and frequencies of
 Ref.~\cite{james} these sums are readily evaluated, and the results are
 shown in Fig.~\ref{sumfig}. From the figure it is seen that
 $\sigma_5,\sigma_6<<\sigma_3,\sigma_4$, so that the last term in 
 Eq.~(\ref{fideldebyeterm}) may be neglected. 
 All the sums have a very rapid convergence and
 we may estimate the fidelity by replacing the sums with their large $N$
 values, {\it i.e.}
\begin{equation}
 F\geq 1-\eta^2 N \frac{\Omega^2}{\nu^2} 0.8 (\bar{n}_1+1)
 \label{fidelothersummed}
\end{equation} 
for the direct coupling (\ref{fidelother}) and 
\begin{eqnarray}
  F&\geq&1-\frac{\pi^2N(N-1)}{8} \eta^4 (1.2 \bar{n}_1^2 + 1.4 \bar{n}_1)  
\label{fideldebyesummed}
\end{eqnarray}
for the Debye-Waller coupling (\ref{fideldebye}).

We note that Eq.~(\ref{fideldebyesummed}) is derived from terms beyond the
Lamb-Dicke expansion and it incorporates the reduction of fidelity due to
deviations from the Lamb-Dicke approximation in the center of mass mode,
{\it cf.} the formal similarity of Eq.~(\ref{fideldebyesummed}) and
Eq.~(\ref{fidelldlowest}). 

\begin{figure}[htbp]
 \begin{center}
   \leavevmode
   \epsfig{angle=270,width=\linewidth,file=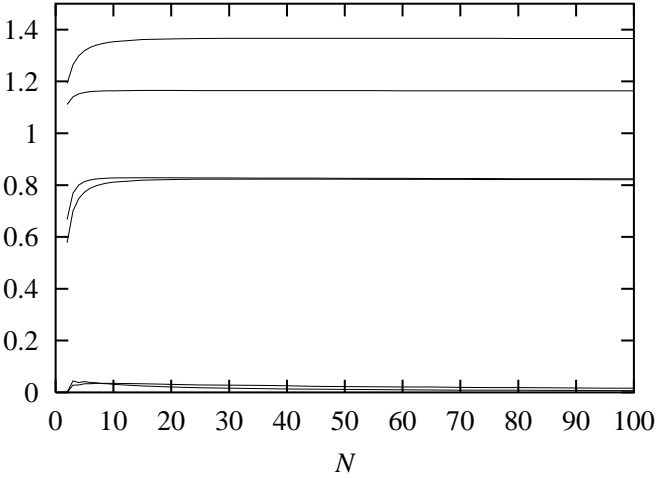}
 \end{center}
 \caption{Evaluation of the sums $\sigma_1...\sigma_6$ for different number
   of ions $N$. Starting from above
   at $N\approx5$
   the curves represent $\sigma_4$, $\sigma_3$, $\sigma_1$, $\sigma_2$,
   $\sigma_6$, and $\sigma_5$. }
 \label{sumfig}
\end{figure}

\subsection{Heating of the vibrational motion}
\label{heating}
An ion trap cannot be perfectly isolated
and the vibration of the ions will be subject to heating due to the
interaction with the environment. Relaxation due to the interaction between
the vibration and a thermal 
reservoir may be  described
by the master equation 
\begin{equation}
  \label{master}
  \frac{d}{dt}\rho = -i {\left[H,\rho\right]}+{\cal L}(\rho),
\end{equation}
where $\cal{L}(\rho)$ is of  the Lindblad form
\begin{equation}
  \label{lindblad}
  {\cal L}(\rho)=-\frac{1}{2}\sum_m {\left(C_m^\dagger C_m \rho +
    \rho C_m^\dagger C_m\right)}+\sum_m C_m \rho C_m^\dagger 
\end{equation}
with relaxation operators $C_1=\sqrt{\Gamma(1+n_{th})}a$ and   $C_2
=\sqrt{\Gamma(n_{th})}a^\dagger$, where $\Gamma$ characaterizes the strength
of the interaction, and $n_{th}$ is the mean vibrational number in thermal
equilibrium. 

We calculate the effect
of heating assuming that the ions remain in the Lamb-Dicke
limit. Changing to the interaction picture with respect to the
Hamiltonian (\ref{xpham}), the time evolution of $\rho$ is entirely due to
the heating, {\it i.e.}, the Lindblad terms which are transformed using the
propagator (\ref{u})   
\begin{eqnarray}
  \label{relax}
  \tilde{C_1}&=&U^\dagger C_1 U=\sqrt{\Gamma(1+n_{th})}{\left(
    a+J_y\frac{G(t)-iF(t)}{\sqrt{2}}\right)}\nonumber\\
  \tilde{C_2}&=&U^\dagger C_2 U=\sqrt{\Gamma n_{th}}{\left(
    a^\dagger+J_y\frac{G(t)+iF(t)}{\sqrt{2}}\right)}.  
\end{eqnarray}

The density matrix is most conveniently expressed in the basis of
$J_y$ eigenstates, and  by tracing over the
vibrational states we find the time derivative of
the internal state density matrix in the interaction picture
\begin{eqnarray}
\label{internalmaster}
\frac{d}{dt} \rho_{M_y,M_y'}&=&-(M_y-M_y')^2\Gamma(1+2n_{th}) \nonumber \\
 &&\quad \quad \times \frac{G(t)^2+ F(t)^2}{4}\rho_{M_y,M_y'}. 
\end{eqnarray}
This equation is readily integrated, and at times $\tau=K2\pi/(\nu-\delta)$
we get
\begin{equation}
  \label{rhoheat}
  \rho_{M_y,M_y'}(\tau)=\rho_{M_y,M_y'}(0){\rm
    e}^{-(M_y-M_y')^2\frac{\Gamma (1+2n_{th})}{4K}\tau}.
\end{equation}
The initial state is expanded on the $J_y$ eigenstates as in
Eq.~(\ref{expansion}) 
and the population of
the initial state (which is ideally constant in the interaction picture)
equals 
\begin{equation}
  \label{fidheat}
  F=\frac{1}{2^{2N}}\sum_{j=0}^N\sum_{k=0}^N  {\left(  
 \begin{array}{cc}
  {N} 
   \\ j
 \end{array} \right)}
 {\left(  
 \begin{array}{cc}
  {N} 
   \\ k
 \end{array} \right)}
  {\rm e}^{-(j-k)^2\frac{\Gamma (1+2n_{th})}{4K}\tau}.
\end{equation}
For two ions this expressions can be readily evaluated
\begin{equation}
  \label{fidheat2}
  F=\frac{3}{8}+\frac{1}{2}   {\rm e}^{-\frac{\Gamma (1+2n_{th})}{4K}\tau}
  +\frac{1}{8}  {\rm e}^{-4\frac{\Gamma (1+2n_{th})}{4K}\tau}.
\end{equation}
In the limit of many ions ($N>>1$) and short times ($\frac{\Gamma
(1+2n_{th})}{4K}\tau<<1$) we may again approximate the expression in
Eq.~(\ref{fidheat}) by assuming that $j$ and $k$ are continuous variables
and by replacing the binomial coefficients by Gaussian distributions with the
same width. In this limit the fidelity becomes
\begin{equation}
  \label{fidheatmany}
  F={{1}\over{\sqrt{1+N\frac{\Gamma (1+2n_{th})}{4K}\tau}}}.
\end{equation}
For 2 ions the deviation between (\ref{fidheat2}) and
(\ref{fidheatmany}) is less than 0.02 for all values of $F$ larger than $0.5$.

In the above expressions we have assumed the Lamb-Dicke approximation. This
corresponds to a situation where the heating is counteracted for example by 
lasercooling on some ions reserved for this purpose. If the ions are not cooled
the heating will proceed towards high vibrational numbers   with a heating
rate  $\Gamma n_{th}$ and the heating will
eventually take the ions out of the Lamb-Dicke limit.  With strong fields
($K\sim 1$)  the reduction in
the fidelity described by Eq.~(\ref{fidheatmany}) will ruin the entangled state
before the heating has made a significant change to the vibrational state
($\Gamma n_{th} \tau \gtrsim 1$). For weak fields ($K>>1$) however, the
situation is different. With weak fields one may produce an entangled state
even though the time required to entangle the ions is much longer than
the decoherence time of the vibrational motion which is used to communicate
between the ions, {\it i.e.} if $K>N\Gamma
n_{th} \tau$ the effect of heating is small even though the change in
the average vibrational number $\Gamma n_{th}
\tau$ is larger than unity
\cite{anders1,anders2}.  Since
the effective Rabi-frequency has a small dependence on the vibrational
quantum number $n$ as described in Eq.~(\ref{omegan}) the heating will have
an indirect effect on the internal state preparation. This can be modelled
by changing the probabilities in Eqs.~(\ref{fidelld}-\ref{fidelldmanygauss})
into time 
dependent functions $P_n(t)$ reflecting the change in the vibrational
motion occurring during the internal state preparation.

\noindent
\setlength{\extrarowheight}{9 pt}
\begin{table*}[bt]
\caption{Creation of entangled states of $N$ ions
  $\frac{1}{\sqrt{2}}(|gg...g\ra -i |ee...e\ra)$ by interaction with a
  bichromatic field (\ref{interaction})   $H_{\rm int}=2\Omega J_x\cos \delta t
  -\sqrt{2}\eta\Omega J_y 
     [  x (\cos(\nu-\delta)t+\cos(\nu+\delta)t)
   +p
  (\sin(\nu-\delta)t+\sin(\nu+\delta)t) ]$ obeying
  $\frac{\eta\Omega}{\nu-\delta}=\frac{1}{2\sqrt{K}}$, $K=1,2,3,...$ and
  for a duration $\tau=2\pi K/(\nu-\delta)$. The fidelity of the preparation
  is reduced by various causes, listed in the table.}
\label{tabel}

\begin{tabular}{m{2.5cm} m{2.5cm} m{3cm} m{3cm} m{3cm} m{3cm}}
\hline\hline
\\
\multicolumn{1}{m{2.5cm}}{\multirow{2}{2.5cm}{Cause of deviation}}   
  &\multicolumn{1}{m{2.5cm}}{\multirow{2}{2.5cm}{Direct off-resonant
      \newline coupling \newline $J_x$ term in (\ref{interaction})}}
  &\multicolumn{1}{m{2.6cm}}{\multirow{2}{2.6cm}{Deviations from Lamb-Dicke
    \makebox[\linewidth]{$\langle n|{\rm
        e}^{i\eta(a+a^\dagger)}|n+1\rangle$} \newline 
    \makebox[\linewidth]{\hspace{1cm}$\neq i \eta
    \sqrt{n+1}$}} } 
  &\multicolumn{2}{c}{ Spectator vibrational modes}
  &\multicolumn{1}{m{3cm}}{\multirow{2}{3cm}{ Heating of the vibration
      towards vibrational  number $n_{th}$ with rate $\Gamma n_{th}$ }}
\\  
  & 
  & 
  & (i) Direct coupling\newline to other modes 
  & (ii) Debye-Waller
  &
\\ 
 $1-F$ 
   & \centering ${N \Omega^2 \over 2 \delta^2}$
   & \centering $\eta^4  {\pi^2 N(N-1) \over 8} {\rm Var}(n_1)$ 
   & \centering $N{\eta^2\Omega^2 \over \nu^2}0.8(
     \bar{n}_1+1)$
   & $\eta^4\frac{\pi^2N(N-1)}{8}$ 
     \makebox[\linewidth]{\hspace{0.0cm} $\times(0.2\bar{n}_1^2 
      +0.4\bar{n}_1)$}
   & \makebox[\linewidth]{\centering $N {\Gamma(1+2n_{th}) \tau \over 8
       K}$}\\
Eq.
   &\centering (\ref{fidelc})
   &\centering (\ref{fidelldlowest})
   &\centering (\ref{fidelothersummed})
   &\centering (\ref{fideldebyesummed}) minus (\ref{fidelldlowest})
   &\makebox[\linewidth]{\centering (\ref{fidheatmany})}   \\
\hline \hline
\end{tabular}
\end{table*}

\section{Conclusion}
\label{conclusion}
We have in this paper evaluated the possibility for preparation of
entangled states of ions by illumination with bichromatic light. We have
identified two regimes: (i) a weak field regime where single photon
absorption is suppressed and where two-photon processes interfere in a way
that makes the internal state dynamics insensitive to the vibrational state,
and (ii) a strong field regime where the individual ions are coherently
excited and the motional state is highly entangled with the internal state
until all undesirable excitations are deterministically removed towards the
end of the interaction. 

We have presented analytical estimates for the fidelity of the internal state
preparation.
These expressions are summarized in table~\ref{tabel}.
The expressions for the fidelity may be readily applied to experimental
parameters and they show that several ion trap experiments today are in a
position to apply our proposal directly. In fact, using our proposal the
NIST group at Boulder has been able to produce the maximally entangled
state $\frac{1}{\sqrt{2}}(|gggg\ra-i|eeee\ra)$ with four ions
\cite{nature}. In this experiment the heating of the center-of-mass mode
was so strong that this mode could not be used to communicate between the
ions. Instead the experiment used an asymmetric mode where all ions have
the same amplitude but a different sign, {\it i.e.} $|\eta_i|$ are the
same for all ions $i$. Apart from the center-of-mode such modes only exist in
ion traps containing two or four ions, and the experiment could not go
beyound four ions. In other existing traps the heating is much less
significant \cite{roos}, and these traps may be employed to produce
entangled states with more particles.

The use of ancillary degrees of fredom (center-of-mass position and
momentum) to communicate between two or more quantum systems is a key
ingredient of quantum information processing. The algebraic property
(\ref{jzsq}) which allows coupling
and temporary entanglement with such an ancilla may find wide
applications in many different systems for quantum computation with
different ancillae  (photons,
phonons, Cooper-pairs, etc.). However,
operators with a constant non-vanishing commutator (which allows the
formal step from Eq.~(\ref{baker}) to Eq.~(\ref{jzsq})) only exist in 
infinite-dimensional Hilbert spaces \cite{commutator}. In addition to the
implementation in cavity QED realizations of quantum computing
\cite{domokos,turchette,imamoglu} where quantized cavity fields play the
role of the vibrational modes, it thus seems very
relevant to investigate to 
which extent the ideas underlying Eq.~(\ref{jzsq}) can be generalized to 
ancillae with a 
finite number of states and, {\it e.g.}, for communication across a linear
qubit register by
only nearest neighbour interaction.

\section*{Acknowledgments}
We thank   B.~E.~King, C.~Monroe, D.~J.~Wineland, R.~Blatt, D.~Leibfried,
and F.~Schmidt-Kahler for  
fruitfull discussions and for enlightening us on details of their trapping
experiments. We also thank D.~F.~V.~James for providing the
eigenfrequencies and modes for the collective vibrations which were used to
produce Fig.~\ref{sumfig}.  This work is  supported by Thomas B. Thriges Center
for Kvanteinformatik and by the Danish National Research Council.


\begin{thebibliography}{99}
\bibitem{commutator} If operators $A$ and $B$ have a constant
  commutator $c\cdot I$, then ${\rm Tr}([A,B])=c\cdot {\rm dim}({\cal
    H})$. If ${\rm dim}({\cal H})$ is finite, the left hand side vanishes due
  to the cyclic property of the trace and thus $c=0$.
\end{thebibliography}
\end{document}